\documentclass[11pt]{article}
\begin{document}
\begin{flushright}

FTUV/02-1107~,~IFIC/02-54 \\
November,7 2002
\vskip .5cm
\end{flushright}

\begin{center}
{\bf\Large On dynamical supergravity interacting 
with super-p-brane sources}\footnote{Invited talk delivered
at the {\it 3rd International Sakharov Conference on
Physics}, Moscow, June 24-29, 2002.
To appear in the Proceedings.}
\vskip .5cm

{\large Igor A. Bandos$^{1,2}$, Jos\'e A. de 
Azc\'arraga$^{1}$, Jos\'e~M.~Izquierdo$^{3}$ and 
Jerzy Lukierski$^4$} 
\vskip .5cm

{\small\it $^1$ Departamento de F\'{\i}sica Te\'orica and IFIC,
46100-Burjassot (Valencia), Spain,\\
$^2$ Institute for Theoretical Physics, NSC KIPT, 
UA61108, Kharkov, Ukraine,\\
$^3$ Departamento de F\'{\i}sica Te\'orica,
Facultad de Ciencias, 47011-Valladolid, Spain,\\
$^4$ Institute for Theoretical Physics,
pl. Maxa Borna 9, 50-204 Wroclaw, Poland}

\end{center}

\centerline{\bf Abstract}
{\small We review recent progress in a fully dynamical Lagrangian 
description of the supergravity--superbrane interaction. It suggests 
that the interacting superfield action, when it exists, is gauge 
equivalent to the component action of dynamical supergravity 
interacting with the bosonic limit of the superbrane.} 

\vskip 1.5cm

\section{Introduction}

Superbranes \cite{BST87,AETW}
play an important r\^{o}le in recent  developments of 
String/M-theory \cite{Mth} as well as in their applications to the 
study of quantum Yang--Mills theories \cite{Maldacena} and the 
structure of the Universe. 

A popular description of superbranes was proposed in 1989 
\cite{DGHR,SUGRA}. It identified them with solitonic 
solutions of the pure bosonic `limit' of the supergravity equations. 
Although all the fermions are set equal to zero, the superbrane solutions are 
supersymmetric and, hence, stable. Already in 
\cite{DGHR}, the superstring description as a solitonic solution of 
$D=10$ dimensional supergravity was found. 
On the other hand, linearized supergravity was derived from the 
quantization of the superstring in flat superspace \cite{GSW}. This allows 
one to 
identify supergravity with the low energy limit of superstring theory. 
As the complete supergravity action 
can be uniquely restored from the linearized one by  
requiring $D=10$, type IIB local supersymmetry, 
the  identification of supergravity as a low energy limit of 
superstring theory is quite convincing although 
the mechanism for the appearance of curved spacetime from the superstring 
is still obscure and requires further progress in constructing string 
field theory \cite{StrF}. 

One may, however, consider the interaction of a superstring with a 
supergravity {\sl background} {\it i.e.}, consider a superstring moving 
in a given curved superspace \cite{strKsg}.  
Then, selfconsistency requires having a smooth flat superspace limit 
for such a system. This implies, in particular, that 
the Green--Schwarz superstring has to  possess local fermionic 
$\kappa$--symmetry \cite{ALS,GSW} in curved superspace, as it does in the flat 
one. Such a requirement immediately results in $D=10$ superfield supergravity 
constraints being imposed on the background superfields \cite{strKsg}. 
The point is that these are {\sl on-shell} constraints, {\it i.e.} 
that their selfconsistency implies equations of motion for 
$D=10$ supergravity (actually, sourceless or `free').

This could be regarded as an advantage of superstring theory, as in the 
bosonic string model the equations of motion for the background can be 
obtained only by one--loop quantum calculations (namely, from the requirement 
that the conformal (Weyl) symmetry is preserved in the quantum theory). 
However, on the other hand, these equations of motion are {\sl sourceless}  
supergravity equations. Then  their appearance from the selfconsistency 
conditions  for the superstring might produce the impression that 
supersymmetry forbids the interaction of supergravity with a source. 

The same situation occurs with other superbranes \cite{BST87,ST90}:  
as their worldvolume actions possess $\kappa$--symmetry when considered 
in the flat superspace, the 
selfconsistency of the model in curved superspace requires a smooth 
flat superspace limit and, hence, the preservation of the 
$\kappa$--symmetry in a curved background. Such a requirement results in 
the supergravity constraints, which again imply `free' supergravity 
equations of motion for 
the physically most interesting $D=10$ and $11$ cases.  

So a challenge is to see whether one can provide a full (quasi)classical 
Lagrangian description of the supergravity--superbrane interacting system 
so that supergravity is treated dynamically and not just given by a 
background. 
Such a Lagrangian should produce a complete set of 
supergravity equations with singular sources having support on 
surfaces (worldvolumes) in spacetime (superspace). 
In this contribution we review briefly 
our recent  results \cite{BAIL,BdAI1,BAIL4} 
in this direction, {\it i.e.} towards a selfconsistent Lagrangian 
description of supergravity--superbrane coupled system.

\section{Statement of the problem}

It is natural to assume that such a Lagrangian description 
is based on the sum of {\sl some} supergravity action  
$S_{SG}$ and the super--$p$--brane action $S_p$, 
\begin{equation}
  \label{S=S+S}
  S= S_{SG} + S_{p}  \; . 
\end{equation}

A superbrane is a brane moving in superspace. 
This means that the super--$p$--brane worldvolume ${\cal W}^{p+1}$ is a 
hypersurface $Z^M= \hat{Z}^M (\xi)$ 
in superspace $\Sigma^{(D|n)}$ with $D$ bosonic coordinates 
$x^\mu$ ($\mu = 0,1,\ldots, D-1$)  
and $n=N$dim$(Spin(1,D-1))$ fermionic coordinates 
$\theta^{\check{\alpha}}$ ($\check{\alpha}=1,\ldots, n$; below
we assume $N$=1 for simplicity), 
\begin{equation}
  \label{cWp}
  {\cal W}^{p+1} \; \subset \; \Sigma^{(D|n)}\; : \
Z^M= \hat{Z}^M (\xi) \; \leftrightarrow 
\cases{x^\mu = \hat{x}^\mu (\xi) \;  \cr 
\theta^{\check{\alpha}}=  \hat{\theta}^{\check{\alpha}} (\xi)}
\; ;   \end{equation}
the $\xi^m = (\tau ,\vec{\sigma})$ are ${\cal W}^{p+1}$ 
local coordinates,  
$m=0, \ldots, p$, and the hat indicates dependence on $\xi$.  
The action  $S_p$ is formulated in terms of supervielbein and 
gauge field superforms on $\Sigma^{(D|n)}$,  
characteristic of the superfield description of supergravity, 
 \begin{eqnarray}
  \label{EA}
& E^A (Z)= dZ^M E_M^A (Z)=(E^a, E^\alpha )\; , 
\\ \nonumber & C_q(Z) = {1\over q!} dZ^{M_q} \wedge 
\ldots \wedge dZ^{M_1} C_{M_1\ldots M_q}(Z)\; \; ,
\end{eqnarray}
pulled back to ${\cal W}^{p+1}$ 
 \begin{eqnarray}
  \label{Sp}
& S_p= \int\limits^{}_{{\cal W}^{p+1}} 
{\cal L}_{p+1} (\hat{E}^a, \hat{C}_{q}) 
\equiv S_p[\hat{E}^a, \hat{C}_q] , 
\\ 
  \label{hEA}
& \hat{E}^A  = E^A (\hat{Z}(\xi))
:= d\hat{Z}^M(\xi) E_M^A(\hat{Z}(\xi))
\; ,  \quad 
\hat{C}_{q}= C_q(\hat{Z}(\xi))
\end{eqnarray}
and, perhaps, a set of worldvolume gauge fields 
(one--form $A(\xi)=d\xi^m A_m(\xi)$ for Dirichlet superbranes). 
For instance, for the 
`standard' branes which do not carry additional worldvolume gauge 
fields, the action $S_p$ reads \cite{AETW}
 \begin{eqnarray}
  \label{Sps}
& S_p = S_p[\hat{E}^a, \hat{C}_{p+1}]= 
\int\limits^{}_{{\cal W}^{p+1}} 
({1\over 2(p+1)}\;* \hat{E}_a \wedge \hat{E}^a - 
\hat{C}_{p+1})\; , \quad 
\end{eqnarray}
where $*$ is the Hodge star operator for the induced metric, 
$* {1\over (p+1)} \hat{E}_{{a}} \wedge \hat{E}^{{a}}=   
d^{p+1}\xi \; \sqrt{|det (\hat{E}_{ma} \hat{E}_n^a)|}$.   

Hence, to have a well posed variational problem for the 
interacting system action (\ref{S=S+S}), one has to assume 
that the supergravity action is also formulated in 
terms of superfields $E_M^A(Z)$, $C_{M_1\ldots M_q}(Z)$, 
\begin{equation}
  \label{S=sf} 
  S= S_{SG}[E^A, C_q]  + S_{p}[\hat{E}^a, \hat{C}_q]  \; . 
\end{equation} 
The problem, however, is that for the most interesting cases 
($D=10$ and $D=11$) 
no {\sl superfield} action for supergravity is known. 

As, in contrast, the {\sl component} action  
$S_{SG}[e^a, e^{\alpha}=dx^\mu \psi_\mu^{\alpha},
\ldots]$
for all $D\leq 11$ supergravity theories  is now known,  
one might think of using it in (\ref{S=sf})
 instead of the superfield supergravity action,  
and decompose the superfields in $S_{p}[\hat{E}^a, \hat{C}_q]$ 
in terms of component fields, {\it e.g.} 
$E^a_\mu= e^a_\mu(x) + {\cal O}(\theta )$, $E_\mu^{\alpha}=  
\psi_\mu^{\alpha}(x)+ {\cal O}(\theta) $. 
However, the problem is that to find  
the  decompositions of these 
superfields one has to use the superspace constraints.
For $D=10, 11$ these are the {\sl on--shell} supergravity 
constraints, which imply that the fields $e^a_\mu(x)$, 
$\psi_\mu^{\alpha}(x)$ obey the {\sl `free'} supergravity 
equations (without any source term). Moreover, as these 
fields are solutions of equations of motion, one is not free 
to take their arbitrary variations. Thus, despite the first 
impression,  the variational 
problem for the action 
$S_{SG}[e^a, dx^\mu \psi_\mu^{\alpha},\ldots ]
+ S_{p}[\hat{E}^a, \hat{C}_q]$ is not well posed.

\section{Group manifold approach to supergravity--superbrane 
interaction} 

A solution 
to the above problem (for any $D$)  
was proposed in \cite{BAIL}
using in (\ref{S=sf})
the {\sl group manifold} or rheonomic action for supergravity
\cite{rheo}, 
\begin{equation}
  \label{SgmSG} 
S_{gmSG}= S_{gmSG}[\tilde{E}^A,\tilde{w}^{ab}, \tilde{C}_q] \; ,  
\end{equation} 
which is written in terms of superfields, but pulled back to a 
surface ${\cal M}^D$ of maximal bosonic dimension in superspace, 
\begin{eqnarray}
  \label{cMD}
  {\cal M}^{D} \; \subset \; \Sigma^{(D|n)} \; :  \; 
Z^M= \tilde{Z}^M (x) \; \Leftrightarrow 
\cases{x^\mu - arbitrary \;  \cr 
\theta^{\check{\alpha}}=  \tilde{\theta}^{\check{\alpha}} (x)}
\; ,   \\ 
  \label{tEtC}
\tilde{E}^A := E^A (\tilde{Z}(x))= d\tilde{Z}^M(x) E_M^A (\tilde{Z}(x))\; , 
\qquad \nonumber \\ 
\tilde{w}^{ab} := w^{ab} (\tilde{Z}(x))\; , \qquad  
\tilde{C}_q := C_q (\tilde{Z}(x))\; , \qquad 
\end{eqnarray} 
where the tilde indicates $x$--dependence. 

As both $S_{gmSG}$ and $S_p$ are formulated in terms of pull--backs of the 
same superfields, the variational problem for the coupled action 
\begin{equation}
  \label{Sgm+Sp} 
S = S_{gmSG}[\tilde{E}^A,\tilde{w}^{ab}, \tilde{C}_q] 
+ S_{p}[\hat{E}^a, \hat{C}_q] 
\;  
\end{equation} 
is now well posed. 

A salient feature of (\ref{Sgm+Sp}) is that the interacting 
dynamical system involves two types of fermionic coordinate functions, 
$\tilde{\theta}^{\check{\alpha}} (x)$ and 
$\hat{\theta}^{\check{\alpha}} (\xi)$, which are apparently 
independent. However, it may be seen \cite{BAIL} that 
the variation with respect to the bosonic supervielbein produces 
an equation the consequences of which include not 
only the superfield (superform) generalization of the Einstein equation  
(with a source)   
pulled back to the surface ${\cal M}^D$, 
\begin{equation}
  \label{MD=J} 
\tilde{M}_{(D-1)a}:= \tilde{R}^{bc} \wedge \tilde{E}^{\wedge (D-3)}_{abc} 
+ \ldots = J_{(D-1)a} \; ,  
\end{equation} 
but also the identification of the supergravity and superbrane 
fermionic coordinate functions on ${\cal W}^{p+1}$ \cite{BAIL}, 
\begin{equation}
  \label{th=th} 
\hat{\theta}^{\check{\alpha}} (\xi)= 
\tilde{\theta}^{\check{\alpha}} (\hat{x}(\xi)) \; . 
\end{equation} 
This implies that, on the mass shell, the worldvolume ${\cal W}^{p+1}$ 
lays on the surface ${\cal M}^D$, ${\cal W}^{p+1}\subset {\cal M}^D$. 
The $(D-1)$ current density distribution $J_{D-1}$
in Eq. (\ref{MD=J}) has the form 
\begin{equation}\label{JD=} 
J^a_{(D-1)}  = dx^{\wedge (D-1)}_\mu \int_{{\cal W}^{p+1}} *\hat{E}^a \wedge 
d\hat{x}^\mu (\tau)  \delta^D(x - \hat{x}(\tau))\; , 
\end{equation} 
and the exterior power is defined by {\it e.g.}, 
$E^{\wedge (D-q)}_{a_1\ldots a_q}= {1\over q!} 
\epsilon_{a_1\ldots a_qb_1\ldots b_{D-q}}E^{b_1}\wedge \ldots 
\wedge E^{b_{D-q}}$ for the bosonic 
supervilebein.

More details on the group--manifold based approach to the 
supergravity--superbrane interacting system can be found 
in ref. \cite{BAIL}. Here we would only like to mention some 
features of this approach which, although reasonable (as seen below 
and in \cite{BdAI1,BAIL4}), were perhaps unexpected.  
\\ {\it i)} 
As the superbrane action (\ref{Sp}) does not contain the pull--back 
$\hat{E}^{\alpha}$ of 
the fermionic supervielbein ${E}^{\alpha}$, 
the variation of the action  (\ref{Sgm+Sp}) with respect to 
${E}^{\alpha}$ produces the pull--back of the superfield (superform) 
generalization of the Rarita--Schwinger equation {\sl without} 
a source term, 
\begin{equation}
  \label{PsiD=0} 
\tilde{\Psi}_{(D-1)\alpha}:= {4i\over 3} {\cal D}\tilde{E}^{\beta}\wedge 
\tilde{E}^{\wedge (D-3)}_{abc} \Gamma^{abc}_{\beta\alpha}  
+ \ldots = 0 \; . 
\end{equation} 
{\it ii)}  
The field equations (Eqs. (\ref{MD=J}), (\ref{PsiD=0}), {\it etc.}) 
are defined 
on an {\sl arbitrary} surface ${\cal M}^D$ in superspace 
(the fermionic function  is not restricted by any equation, except for 
(\ref{th=th}) which could rather be treated as a relation for
$\hat{\theta}^{\check{\alpha}} (\xi)$). Nevertheless, in 
contrast with the case of `free' supergravity, these equations 
cannot be extended (`lifted') to the whole superspace. 
\\ {\it iii)}  The local supersymmetry 
\begin{equation}
  \label{lsusy} 
\delta_{ls} \tilde{E}^a = - 2i \tilde{E}^{\alpha} \Gamma^{a}_{\alpha\beta}
\epsilon^{\beta}(x, \tilde{\theta}(x))\; , \quad 
\delta_{ls} \tilde{E}^{\alpha}= {\cal D} 
\epsilon^{\alpha}(x, \tilde{\theta}(x))+\ldots \;  \quad \ldots 
\end{equation} 
(pull--back of the variational version of the superspace general coordinate 
symmetry on ${\cal M}^D$; $ \delta_{ls} x^\mu =0 = 
\delta_{ls}\tilde{\theta}^{\check\alpha}(x)$, 
{\it cf.} $\tilde{\delta}_{gc}$ in \cite{BdAI1}), 
which is a gauge symmetry of the group manifold action (\ref{SgmSG}), 
is partially ($1/2$) broken on the worldvolume ${\cal W}^{p+1}$ 
for the coupled  system (\ref{Sgm+Sp}). The 
$1/2$ of the supersymmetry preserved on ${\cal W}^{p+1}$ has the form 
of a $\kappa$--symmetry transformation. To formulate the statement in 
a more precise manner, note that 
the superfield $\epsilon^{\alpha}(x, {\theta})$ can be easily restored 
from its pull--back on ${\cal M}^D$, $\tilde{\epsilon}^{\alpha}:=
\epsilon^{\alpha}(x, \tilde{\theta}(x))$. Then this superfield 
parameter of local supersymmetry (\ref{lsusy}) is arbitrary in all 
`points' of  superspace, except for the worldvolume ${\cal W}^{p+1}$, 
where  it  is restricted by the condition 
\begin{equation}
  \label{lsusyE} 
\hat{\epsilon}^{\alpha}:= \epsilon^{\alpha}(\hat{x}(\xi), \hat{\theta}(\xi)) 
= (1-\bar{\gamma})^{\alpha}{}_{\beta}\kappa^{\beta}(\xi) \; . 
\end{equation}
In (\ref{lsusyE})   
 $\bar{\gamma}^{\alpha}{}_{\beta}$ is the kappa--symmetry projector 
(see \cite{AETW}) which satisfies 
$(1-\bar{\gamma})^2= 2(1-\bar{\gamma})$ (as a result, only $1/2$ of the 
components of spinor $\kappa^{\beta}(\tau)$ enter effectively in 
(\ref{lsusyE}) and, hence, in field transformations).
On the other side, $\kappa$--symmetry does not exist
as a separate symmetry of the action (\ref{Sgm+Sp}). 
\\ {\it iv)}  There exists another fermionic gauge symmetry 
(the pull--back of superspace superdiffeomorphism symmetry  
\cite{BdAI1}) 
which transforms the fermionic coordinate function by 
 \begin{equation}
  \label{sdif} 
\delta_{sdiff} \tilde{\theta}^{\check{\alpha}}(x)= 
\varepsilon^{\check{\alpha}}(x, \tilde{\theta}(x))\; .
\end{equation}
This clearly allows one to fix the gauge 
 \begin{equation}
  \label{tthGAUGE} 
\tilde{\theta}^{\check{\alpha}}(x)=  0 \; .
\end{equation}
In the light of the identification (\ref{th=th}), Eq. (\ref{tthGAUGE}) 
implies  also 
\begin{equation}
  \label{hth=0} 
\hat{\theta}^{\check{\alpha}}(\xi)=  0 \; .
\end{equation}
As setting $\tilde{\theta}(x)=0$ in the group manifold action 
(\ref{SgmSG}) one arrives at a first order component action for supergravity 
(written in terms of spacetime fields), 
\begin{equation}
 S_{gmSG}[\tilde{E}^A,\tilde{w}^{ab}, \tilde{C}_q]\vert_{\tilde{\theta}(x)=0}
\equiv S_{SG}[e^a(x), dx^\mu \psi_\mu^\alpha(x) , \omega^{ab}(x)]
\; , 
\end{equation}
one concludes that in the gauge (\ref{tthGAUGE}) the group--manifold 
based action for interacting supergravity--superbrane system, Eq. 
 (\ref{Sgm+Sp}), becomes the action for (component) supergravity 
interacting with a purely {\sl bosonic} brane.  

\section{Dynamical supergravity interacting with a bosonic brane} 

The supergravity interaction with a bosonic brane described by the sum of 
the component action for supergravity and the action for the bosonic brane,  
\begin{equation}\label{SG+bb}
S= 
S_{SG}[e^a(x), dx^\mu \psi_\mu^\alpha(x) , \omega^{ab}(x)] + 
S_p[\hat{e}^a, \hat{C}_{q}]
\;  ,
\end{equation}
could be expected to be not supersymmetric and even inconsistent. However, 
as shown in  \cite{BdAI1}, {\sl the dynamical system  (\ref{SG+bb}) is 
selfconsistent when the bosonic brane is the pure bosonic `limit' of some 
superbrane} (its `associated superbrane') due to the fact 
that in this case {\sl a one--half of the local supersymmetry 
is preserved} on the worldvolume of the bosonic brane. 
To be more precise, when the bosonic brane is the limit of 
a superbrane, the dynamical system 
(\ref{SG+bb}) possesses a local supersymmetry with parameter 
$\epsilon^{\alpha}(x)$  arbitrary out of the worldvolume 
but restricted on the worldvolume by the conditions 
\begin{equation}
  \label{lsusyE0} 
\hat{\epsilon}^{\alpha}:= \epsilon^{\alpha}(\hat{x}(\xi)) 
= (1-\bar{\gamma}_0)^{\alpha}{}_{\beta}\kappa^{\beta}(\xi) \; 
\end{equation}
({\it cf.} Eq. (\ref{lsusyE})), 
where $\bar{\gamma}_0{}^{\alpha}{}_{\beta}= 
\bar{\gamma}^{\alpha}{}_{\beta}\vert_{\hat{\theta}=0}$ is the 
pure bosonic `limit' of the $\kappa$--symmetry projector 
of the associated superbrane.

An inconsistency could now be expected as a result of this loss of  
supersymmetry. Indeed, local supersymmetry is used to gauge away the 
spin 1/2 irreducible representation included in  the gravitino field 
$\psi_\mu^\alpha(x)$. But if local supersymmetry is broken, this 
spin 1/2 part could remain `dynamical' and produce a spin 1/2 ghost. 
Thus, it would appear that there are still problems even if 
the bosonic brane is the bosonic limit of a superbrane, because 
in this case $1/2$ of the supersymmetry is broken on the worldline 
due to 
(\ref{lsusyE0}).  

However, it was shown in \cite{BdAI1} that the presence of a source 
in the Einstein equation (the $\tilde{\theta}(x)=\hat{\theta}(\xi)=0$ 
`limit' of 
Eq. (\ref{MD=J})) and the absence of a source in the 
gravitino equation ($\tilde{\theta}(x)=\hat{\theta}(\xi)=0$ `limit' of 
(\ref{PsiD=0})) results in a condition on the supercurrent which, in  turn, 
is equivalent to a {\sl fermionic equation for the bosonic brane} variables 
which include the pull--back $\hat{e}^{\alpha}$ of the fermionic form 
$e^\alpha=dx^\mu \psi_\mu^\alpha$  
(specifically, $\hat{e}^{\beta}\, 
\Gamma^a_{{\alpha}{\beta}}\, \hat{e}_{\tau a} =0$
for a bosonic particle; this 
is the bosonic `limit' of the superparticle equation 
$\hat{E}^{\beta}\, 
\Gamma^a_{{\alpha}{\beta}}\, \hat{E}_{\tau a} =0$). 
As shown in \cite{BdAI1}, precisely these {\sl fermionic equations 
for the bosonic brane} take up the r\^ole of gauge fixing conditions for 
the part of 
local supersymmetry that is lost on the bosonic brane worldvolume 
due to  (\ref{lsusyE0}). In other words, $\hat{e}^{\beta}\, 
\Gamma^a_{{\alpha}{\beta}}\, \hat{e}_{\tau a} =0$ or its higher $p$
counterpart, together with the remaining $1/2$ of the 
local supersymmetry, remove the spin $1/2$ ghost 
part of the gravitino field on the worldvolume of the bosonic brane.

To clarify further the reasons  for such selfconsistency of the 
supergravity--bosonic brane system, we have investigated in \cite{BAIL4} 
a full superfield description of the supergravity--superparticle 
interacting system in a simple case where the superfield supergravity 
action exists, namely in $D=4$ $N=1$ superspace, to which we now turn.

\section{Superfield description of the $D=4$, $N=1$ 
supergravity--superparticle interacting system}

The superfield action for supergravity--superparticle 
interacting system is  
\begin{equation}  \label{S=sf0} 
  S= S_{SG\, 4D}[E^A] + S_{0}[\hat{E}^a]  \; , 
\end{equation}
where 
\begin{eqnarray}\label{SGWZ}
S_{SG\, 4D}[E^A]= 
\int d^4 x {d}^4 \theta \; sdet(E_M^A) \; 
\equiv \int d^8 Z  \; E \; 
\end{eqnarray} 
is the Wess--Zumino superfield action for  $D=4$, $N=1$ supergravity 
\cite{WZ77}, and 
  \begin{eqnarray}
  \label{Sp0}
S_0:= S_{0}[\hat{E}^a]= {1\over 2}
\int\limits^{}_{{\cal W}^{1}} l(\tau) \hat{E}_{\tau a} \hat{E}^a  \;  
\end{eqnarray}
is the Brink--Schwarz superparticle action ($\hat{E}^a = 
d\hat{Z}^M(\tau) E^a_M(\hat{Z})= 
d\tau \hat{E}^a_{\tau}$ and $l(\tau)$ is a Lagrange multiplier 
which can be treated as a worldline einbein).

In Eq. (\ref{SGWZ}) the supervielbein $E_M^A(Z)$ is  
assumed to obey the supergravity constraints
$T_{\alpha \dot{\beta}}{}^a= -2i\sigma^a_{\alpha \dot{\beta}}$
as well as $T_{\alpha\beta}{}^A=0=T_{\dot{\alpha}\dot{\beta}}{}^A$, 
$T_{\alpha\dot{\beta}}{}^{\dot{\gamma}}=0$,  $T_{\alpha b}{}^c=0$, 
and $R_{\alpha\dot{\beta}}{}^{ab}=0$ (or $T_{ab}{}^c=0$ as, {\it e.g.},  
in \cite{BW}).
This makes the variational problem slightly subtle (see \cite{WZ77,BAIL4}), 
but nevertheless well posed. 

The superfield equations of motion 
which result from the variation of the coupled action (\ref{S=sf0}) 
with respect to the superfield variables turn out to be 
(see  \cite{BAIL4})
\begin{eqnarray}
\label{SGeqmR}
\bar{R}:= -{1\over 12} R_{{\alpha}{\beta}}{}^{ab} 
(\sigma_a \tilde{\sigma}_b)^{{\alpha}{\beta}}=0 \; , 
\\ 
\label{Ga=Ja}
 G_a:= 2i (T_{a\beta}{}^\beta -T_{a\dot{\beta}}{}^{\dot{\beta}})
= {\cal J}_a 
\; ,  
\end{eqnarray} 
which imply the following superfield generalization of the 
Rarita--Schwinger and Einstein   equations for the coupled system 
\begin{eqnarray}\label{SGRS=J1}
& \Psi^a_{\dot\alpha}  \equiv 
\epsilon^{abcd}T_{bc}{}^{\alpha}\sigma_{d\alpha\dot{\alpha}}= 
{i\over 4} \bar{{\cal D}}_{\dot{\alpha}}
{\cal J}^a   \; ,
\\ \label{EiInt1} 
& R_{bc}{}^{ac} = {1\over 16} 
\tilde{\sigma}_b^{\dot{\beta}\beta} [{\cal D}_{\beta}, 
\bar{{\cal D}}_{\dot{\beta}}] {\cal J}^a \; .
\end{eqnarray}

The current potential ${\cal J}_a$ entering Eq. (\ref{Ga=Ja}) (and 
Eqs. (\ref{SGRS=J1}), (\ref{EiInt1})) 
is expressed through 
the set of current prepotentials 
\begin{eqnarray}
\label{K2,3/2}
& {\cal K}_a^{B}(Z):= \int_{{\cal W}^1}  
 {l(\tau )\over \hat{E}} \hat{E}_{a\tau} \hat{E}^{B} 
\delta^8(Z-\hat{Z})  \, , \quad \delta^8(Z-\hat{Z}) := \delta^4(x-\hat{x}) 
(\theta-\hat{\theta})^4 \, , \quad 
\end{eqnarray}
by  
\begin{eqnarray} 
\label{Ja=}
& {1\over 6}{\cal J}_a = - i {\cal D}_{\alpha} {\cal K}_a^{\alpha} 
+ i \bar{{\cal D}}_{\dot{\alpha}} \bar{{\cal K}}_a^{\dot{\alpha}} 
+ {1\over 4} \tilde{\sigma}_b^{ \dot{\alpha} {\alpha} }
[{\cal D}_{{\alpha}}, \bar{{\cal D}}_{\dot{\alpha}}] {\cal K}_a{}^b \; .
\end{eqnarray}
and, {\sl due to the structure of the supergravity equations},  
satisfies the identities 
 \begin{eqnarray}\label{DJa=0}
{\cal D}^{\alpha} {\cal J}_{\alpha\dot{\alpha}} = 0 \; , \qquad 
\bar{{\cal D}}^{\dot{\alpha}}{\cal J}_{\alpha\dot{\alpha}} = 0 \; , 
\qquad  {\cal D}^{a} {\cal J}_{a} = 0 \quad 
\end{eqnarray}
($\sigma_{\alpha\dot{\alpha}}^a {\cal J}_{a}
\equiv{\cal J}_{\alpha\dot{\alpha}}$)
{\sl which are equivalent to the superparticle equations of motion}, 
\begin{eqnarray}\label{fEqm}
&\hat{E}^\alpha  \sigma^a_{\alpha\dot{\alpha}} \hat{E}_{a\tau} =0 \; , \qquad 
\hat{E}_{a\tau} \sigma^a_{\alpha\dot{\alpha}} \hat{\bar{E}}{}^{\dot{\alpha}} 
= 0 \; ,  \qquad  
 {\cal D}(l\, \hat{E}_{a\tau}) = 0\; .  
\end{eqnarray}
(only the superparticle Lagrange multiplier $l(\tau)$ produces an independent
  equation, $\hat{E}^a_{\tau} \hat{E}_{a\tau}=0$). 
Note that the bosonic counterpart of the above statement is well known in 
general relativity (see, {\it e.g.} 
pp. 19, 44-48 and Eq. (1.6.13) in  \cite{Inf} and p. 240 in \cite{Fock}).

The above mentioned dependence of 
the superparticle equations of motion is 
 the content of a Noether identity which, 
in the language of the second Noether theorem (see \cite{BdAI1,BdAI2}), 
reflects the fact that the action (\ref{S=sf0}) possesses a gauge symmetry 
(superspace superdiffeomorphisms) 
the transformations of which act additively on the superparticle coordinate 
functions. In other words, the superparticle coordinate functions behave 
like {\sl compensators} (pure gauge fields) with respect to 
superspace diffeomorphism symmetry. 

This is tantamount to saying that 
in the supergravity--superparticle interacting system described by the action 
(\ref{S=sf0})
the superparticle coordinate functions $\hat{x}^\mu(\tau)$, 
$\hat{\theta}^{\check{\alpha}}(\tau)$ 
are  the Goldstone fields for a gauge symmetry which is the 
superdiffeomorphism symmetry. The fact that the Goldstone fields are 
defined on 
the worldline or the worldvolume rather than on the whole 
superspace makes the 
super--Higgs effect \cite{Volkov} rather subtle 
in the presence of superbranes (we plan to discuss this separately). 
Here we wish to note only that the Goldstone nature of the 
fermionic coordinate 
functions {\it i.e.}, its additive transformation law under 
superdiffeomorphisms, allows one to fix the fermionic `unitary' gauge 
\begin{eqnarray}\label{thGAUGE}
\hat{\theta}^{\underline{\alpha}}(\tau) = 0 \;  . 
\end{eqnarray}
It was shown in \cite{BAIL4} that in this gauge the current 
potential is proportional to the superspace Grassmann coordinate 
$\theta$ (not to be confused with $\hat{\theta}(\tau)$): 
\begin{eqnarray} \label{Ja=th3}
\hat{\theta}(\tau)=0 \;  
\qquad \Rightarrow \qquad 
{\cal J}_a \propto (\theta)^2 \; . 
\end{eqnarray}
As a result ${\cal D}{\cal J}_a\vert_{\theta =0}=0$ and the 
spacetime gravitino field equation for the supergravity--superparticle 
interacting system (given by the leading component of the 
superfield equation (\ref{SGRS=J1})) becomes sourceless
in the gauge (\ref{thGAUGE}), 
\begin{eqnarray}\label{SGRS=0}
\hat{\theta}(\tau)=0 \;  
\quad \Rightarrow  \quad
\Psi^a_{\dot\alpha} \vert_{\theta=0}\equiv 
2 e e_\mu^a \epsilon^{\mu\nu\rho\sigma} {\cal D}_{[\nu}\psi_{\rho]}^{\alpha}(x)
e^b_\sigma\sigma_{b\alpha\dot{\alpha}}=0 \; .
\end{eqnarray} 
In contrast, the component Einstein equation retains the source term 
in the gauge (\ref{thGAUGE}),  
\begin{eqnarray}\label{EiInt02} 
& e(x) 
R_{bc}{}^{ac} \vert_{\theta=0} = c 
\int_{{\cal W}^1}  
 l(\tau )[ \hat{e}_{b\tau} \hat{e}^{a}] 
\delta^4(x-\hat{x}) \; . 
\end{eqnarray}
The explicit form of the {\it r.h.s}'s of Eqs. (\ref{SGRS=0}), (\ref{EiInt02}) 
uses the Wess--Zumino gauge, which 
can be fixed simultaneously with the gauge  (\ref{thGAUGE}) \cite{BAIL4}.

Eqs. (\ref{SGRS=0}), (\ref{EiInt02}) coincide with the graviton and 
gravitino equations for the supergravity--bosonic particle interacting 
system.  Moreover, one can show (see \cite{BAIL}) that the supergravity 
auxiliary fields vanish (on shell) in this gauge and that,  
in it, the 
superfield action (\ref{S=sf0}) reduces to the action of the 
supergravity--bosonic particle interacting system 
(\ref{SG+bb}) upon integration over the Grassmann coordinates and 
elimination of the supergravity auxiliary fields by 
using  their purely algebraic equations of motion.

Thus, the superfield description of $D=4$ $N=1$ supergravity--superparticle 
interacting system is gauge equivalent to the dynamical system of 
component supergravity interacting with the massless bosonic particle.

\section{Conclusions}

The example of the above 
$D=4$, $N=1$ interacting system shows that the complete 
coupled superfield action possesses all the gauge symmetries characteristic 
of the `free' superfield supergravity and of the superparticle, including 
worldvolume $\kappa$--symmetry. On the other hand, the gauge fixed 
version of the superfield coupled action describes the component 
supergravity interacting with a bosonic particle; moreover, such a gauge 
fixed action produces the gauge fixed version of {\sl all} dynamical 
equations of the superfield interacting system. This suggests that 
the supergravity--superbrane interacting system described 
by a complete {\sl superfield} action (still unknown for the 
interesting cases of $D=10$ and $11$ supergravity) is gauge equivalent 
to the system of dynamical supergravity interacting with the bosonic 
brane obtained by taking the bosonic `limit' of the superbrane. 
 
As the {\sl component} supergravity actions are known 
for all $D\leq 11$  supergravities,  
the above mentioned gauge equivalence would allow one to study any 
supergravity--superbrane interacting system, including systems of 
$D=10, 11$ supergravity interacting with super--$D$--branes and 
super--$M$--branes.  The bosonic brane action appearing as a 
gauge fixed version of the superbrane action keeps trace of its 
origin: when it is added to the (component) action for supergravity, 
the resulting coupled action still possesses  on the worldvolume 
half of the `free' supergravity local supersymmetry.  
This `preserved' part of the local supersymmetry is defined through  
the (pure bosonic limit of the) $\kappa$--symmetry projector 
and is a remnant of the original superbrane $\kappa$--symmetry.

Thus the work outlined here may provide a convenient framework for a
further study of various aspects of brane physics in string/M-theory 
and its applications. In particular, it might be useful  
in the search for new solutions of the supergravity equations with 
nonvanishing fermionic fields and in the analysis of 
anomalies in M-theory (see \cite{LMT}). 
\vskip .2cm

\textbf{Acknowledgments.} 
This work has been partially supported by Spanish Ministry 
of Science and Technology grants BFM2002-03681, 
BFM2002-02000 and EU FEDER funds, the Ucrainian FFR 
(project $\# 383$), INTAS (project N2000-254) and by the 
Polish KBN grant 5P03B05620.


\vskip .5cm

{\small 
\begin{thebibliography}{99}


\bibitem{BST87}
E. Bergshoeff, E. Sezgin and P.K. Townsend, 
%SUPERMEMBRANES AND ELEVEN-DIMENSIONAL SUPERGRAVITY, 
{\em Phys. Lett.} {\bf B189}, 75-78 (1987); {\em Ann. Phys. (NY)} 
{\bf 185}, 330 (1988). 
 
\bibitem{AETW}
A. Ach\'ucarro, J.M. Evans, P.K. Townsend and D.L. Wiltshire, 
%SUPER P-BRANES.
{\em Phys. Lett.} {\bf B198}, 441 (1987).  

\bibitem{Mth}
J.H. Schwarz, 
%{\sl Second Superstring Revolution}, 
{\em Nucl. Phys. Proc. Suppl.} {\bf B55}, 1-32  
(1997) (hep-th/9607067); 
 \, P.K. Townsend, 
{\sl Four Lectures on M--theory}, hep-th/9612121.
%in: {\sl Summer School 
%High energy physics and cosmology}, Trieste 1996, 
%(E. Gava, {\it et al.} Eds.)
%World Scientific, 
%Singapore, 1997. 

\bibitem{Maldacena}
O. Aharony, S.S. Gubser, J. Maldacena, H. Ooguri and Y. Oz, 
%{\sl Large N Field Theories, String Theory and Gravity}, 
{\em Phys. Rep.} {\bf  323},  
183-386, (2000) (hep-th/9905111) and refs. therein.  

\bibitem{DGHR}
 A. Dabholkar, G.W. Gibbons, J.A. Harvey and F. Ruiz Ruiz, 
%  {\sl Superstrings and solitons}, 
{\em Nucl. Phys.} {\bf B340}, 33-55 (1990).  

\bibitem{SUGRA}
M.J. Duff, R.R. Khuri and J.X. Lu, 
% {\sl String Solitons}, 
{\sl Phys. Rep.} {\bf 259}, 213-326 (1995); \\ 
K.S. Stelle, {\sl BPS Branes in Supergravity}, hep-th/9803116.
%in: \textit{High-Energy Physics And Cosmology, Proc.  ICTP Summer School}, 
%2 Jun - 11 Jul 1997, Trieste, Italy   
%(E. Gava, {\it et al.} Eds.)
%%A. Masiero, K.S. Narain, S. Randjbar-Daemi, 
%%G.~Senjanovic, A. Smirnov, Q. Shafi), 
%World Scientific, Singapore, 1998, p. 29-127 
%(hep-th/9803116) {and refs. therein}. 

\bibitem{GSW}
M. Green, J. Schwarz and E. Witten, {\sl Superstring Theory}, I, II, 
CUP, 1987. 

\bibitem{StrF}
E. Witten, 
%{\sl Noncommutative geometry and string field theory},  
{\em Nucl. Phys.} {\bf B268}, 253 (1986); 
W. Siegel, {\sl Introduction to String Field Theory}, 
World Scientific, Singapore, 1988 
(hep-th/0107094); 
I.Ya. Aref'eva, D.M. Belov, A.S. Koshelev and P.B. Medvedev, 
%{\sl 
%Gauge invariance and tachyon condensation in cubic 
%superstring field theory}, 
{\em Nucl. Phys.} {\bf B638}, 21-40 (2002) (hep-th/0107197);  
I. Bars and Y. Matsuo, 
%{\sl Computing in String Field Theory Using the 
%Moyal Star Product}, 
{\em  Phys. Rev.} {\bf D66}, 066003 (2002) 
(hep-th/0204260), and refs. therein.  

\bibitem{strKsg}
M.T. Grisaru, P.S. Howe, L. Mezincescu, B. Nilsson and P.K. Townsend, 
%{\sl Superstring in supergravity background},  
{\em Phys. Lett.} {\bf B162}, 116 (1985).  

\bibitem{ALS}
J. A. de Azc\'{a}rraga and J. Lukierski,
%{\sl Supersymmetric Particles With Internal Symmetries and Central Carges},
{\em Phys. Lett.} {\bf B113}, 170 (1982); 
%SUPERSYMMETRIC PARTICLE MODEL WITH ADDITIONAL BOSONIC COORDINATES, 
{\em Phys. Rev.}  {\bf D28}, 1337 (1983); 
W. Siegel, 
%{\sl Hidden local supersymmetry in the supersymmetric particle action},
{\em Phys. Lett.} {\bf B128}, 397 (1983).   
 
\bibitem{ST90}
J.A. Shapiro and C.C. Taylor, 
%{\sl The spacetime supersymmetric formulation of superstring}, 
 {\em Phys. Rep.} {\bf 191}, 221-287 (1990) and refs. therein. 

\bibitem{BAIL}
I.A. Bandos, J. A. de 
Azc\'arraga,
J. M. Izquierdo and J. Lukierski, 
%{\sl An action for supergravity interacting with super-p-brane sources}, 
{\em Phys. Rev.} {\bf D65}, 021901 (2002) ({hep-th/0104209}). 

\bibitem{BdAI1} 
I.A. Bandos, J.A. de Azc\'arraga and J.M. Izquierdo,  
%{\sl Supergravity interacting with bosonic $p$--branes and local 
%supersymmetry}, 
{\em Phys. Rev.} {\bf D65}, 105010 (2002) (hep-th/0112207). 

\bibitem{BAIL4} 
I.A. Bandos, J. A. de Azc\'arraga, J. M. Izquierdo and J. Lukierski, 
{\sl $D=4$ supergravity dynamically coupled to 
a massless superparticle in a superfield Lagrangian approach}, 
hep-th/0207139. 

\bibitem{rheo}
Y. Neeman and T. Regge, 
{\em Riv. Nuovo Cim.} {\bf 1}, 1 (1978); 
L. Castellani, R. D'Auria and  P. Fr\'e, {\sl Supergravity and
superstrings, a geometric perspective}, vol. 2, World Scientific, 
1991, and references therein. 

\bibitem{WZ77} J. Wess and B. Zumino, 
%SUPERSPACE FORMULATION OF SUPERGRAVITY.
{\em Phys. Lett.} {\bf B66}, 361-364 (1977) 

\bibitem{BW} 
J. Wess and J. Bagger, {\sl Supersymmetry and Supergravity}, 
Princeton University Press, 1992.
%, 259pp. 

\bibitem{BdAI2} 
I.A. Bandos, J.A. de Azc\'arraga and J.M. Izquierdo,  
{\sl On the local symmetries of gravity and supergravity models}, 
hep-th/0201067.
%{\it Preprint} {\bf FTUV/02-1001, IFIC/02-02},
%in: {\it ``Supersymmetries and Quantum Symmetries'' (SQS01)}, 
%Proceedings of XVI Max Born Symposium, 
%Karpacz, Poland, September 21-25, 2001.
%Dubna 2002 (Ed. A. Pashnev), p. 205-221 ({hep-th/0201067}).

\bibitem{Inf}
L. Infeld and J. Plebanski, {\it Motion and Relativity}, 
Pergamon Press, 1960.
%, Warszawa, 229pp. 

\bibitem{Fock}
V. Fock, {\it The theory of space, time and relativity}, Pergamon Press, 
%London, 
1966.
%, 448pp.

\bibitem{Volkov}
D.V. Volkov and V.A. Soroka, {\em JETP Lett.} {\bf 18}, 312 (1973)

\bibitem{LMT}
E. Witten, 
%{\sl Five-Brane Effective Action In M-Theory}, 
{\em J. Geom. Phys.} {\bf  22}, 103 (1997) (hep-th/961023);  
K. Lechner, P.A.~Marchetti and M. Tonin, 
%{\sl Anomaly free effective action for the elementary M5-brane}, 
{\em Phys. Lett.} {\bf B524}, 199-207 (2002) (hep-th/0107061);  
%[arXiv:hep-th/0107061]; \\ 
M. Cariglia and K. Lechner, 
%{\sl NS5-branes in IIA supergravity and gravitational anomalies}, 
{\em Phys. Rev.} {\bf D66}, 045003 (2002) 
(hep-th/0203238) and refs. therein. 

\end{thebibliography}
}
\end{document}